\documentclass[10pt,conference]{IEEEtran}
\usepackage{amsmath}
\usepackage{eqparbox}
\usepackage[usenames, dvipsnames]{color}
\usepackage{steinmetz}
\usepackage{mathtools,amssymb,bm,mathabx,nccmath}
\usepackage{amstext}
\usepackage{amssymb,bm}
\usepackage{graphicx}
\usepackage{color}
\usepackage{booktabs}
\usepackage{longtable}
\usepackage{multicol}
\usepackage{algorithmicx}
\usepackage[ruled]{algorithm}
\usepackage{algpseudocode}
\usepackage{algpascal}
\usepackage{algc}
\usepackage{amsfonts}
\usepackage{dsfont}
\usepackage{array}
\usepackage[most]{tcolorbox}
\usepackage{cases}
\usepackage{pgfplots}
\pgfplotsset{compat=newest}
\usetikzlibrary{spy}
\usepackage{accents}
\usepackage{etoolbox}
\usepackage{caption}
\usepackage{float}
\usepackage{subcaption}
\usepackage{xr-hyper}
\usetikzlibrary{3d}
\usepackage{amsthm}
\DeclareCaptionLabelSeparator{periodspace}{.\quad}
\usetikzlibrary{positioning}
\usepackage{tikz-3dplot}
\usetikzlibrary{matrix}
\usepackage{float}
\usepackage{cite}
\usepackage{epstopdf}
\usepackage[labelfont=bf,font=small]{caption}
\makeatletter
\renewcommand{\fnum@figure}{Fig. \thefigure}
\makeatother
\usepackage{etoolbox}
\theoremstyle{remark}

\newcommand\ASTART{\bigskip\noindent\begin{minipage}[b]{0.5\linewidth}}
	
	\newcommand\AENDSKIP{\end{minipage}\bigskip}
\newcommand\AEND{\end{minipage}}
\ifCLASSOPTIONcaptionsoff
\usepackage[nomarkers]{endfloat}
\let\MYoriglatexcaption\caption
\renewcommand{\caption}[2][\relax]{\MYoriglatexcaption[#2]{#2}}
\fi
\theoremstyle{plain}

\newtheorem{prop}{\textbf{Proposition}}

\theoremstyle{definition}

\newtheorem{rem}{\textbf{Remark}}

\newcommand*{\rom}[1]{\expandafter\@slowromancap\romannumeral #1@}

\newcommand{\RN}[1]{%
\textup{\uppercase\expandafter{\romannumeral#1}}%
}

\usepackage{standalone}
\graphicspath{{./Figures/}}
\usepackage{times}
\usepackage[T1]{fontenc}
\usepackage[english]{babel}
\usepackage{graphicx}
\usepackage{amsmath, amsfonts, dsfont,amssymb, graphicx, array, tabularx, booktabs,multicol}
\usepackage{amsthm}
\usepackage{epsf,epsfig}
\usepackage{setspace}

\usepackage{lipsum}
\usepackage{mathrsfs}
\usepackage{multirow}
\usepackage{url}
\usepackage{color}
\usepackage[nolist,printonlyused]{acronym}      
\usepackage{comment}
\usepackage{cite}
\usepackage{bm}
\usepackage{tikz}
\usetikzlibrary{angles,quotes}
\usetikzlibrary{decorations.pathreplacing}
\usepackage[labelfont=bf,font=footnotesize]{caption}

\usepackage{mathtools}
\usepackage{xfrac}
\usepackage{blindtext}

\usepackage{stfloats}

\newtheorem{remark}{Remark}

\newcommand{\mx}[1]{\mathbf{#1}}

\definecolor{amber}{rgb}{1.0, 0.49, 0.0}
\definecolor{ao}{rgb}{0.0, 0.5, 0.0}

\def\R2#1{\textcolor{black}{#1}}
\def\R3#1{\textcolor{black}{#1}}

\definecolor{copperrose}{rgb}{0.6, 0.4, 0.4}
\definecolor{azure}{rgb}{0.0, 0.5, 1.0}
\definecolor{ashgrey}{rgb}{0.7, 0.75, 0.71}
\definecolor{chestnut}{rgb}{0.8, 0.36, 0.36}
\definecolor{airforceblue}{rgb}{0.36, 0.54, 0.66}
\definecolor{cadmiumorange}{rgb}{0.93, 0.53, 0.18}
\definecolor{bleudefrance}{rgb}{0.19, 0.55, 0.91}
\definecolor{carolinablue}{rgb}{0.6, 0.73, 0.89}
\definecolor{blue(ncs)}{rgb}{0.0, 0.53, 0.74}
\definecolor{dodgerblue}{rgb}{0.12, 0.56, 1.0}
\definecolor{cssgreen}{rgb}{0.0, 0.5, 0.0}
\definecolor{cadmiumgreen}{rgb}{0.0, 0.42, 0.24}
\definecolor{cadmiumorange}{rgb}{0.93, 0.53, 0.18}
\definecolor{amaranth}{rgb}{0.9, 0.17, 0.31}
\definecolor{bluegray}{rgb}{0.4, 0.6, 0.8}
\definecolor{cerulean}{rgb}{0.0, 0.48, 0.65}
\definecolor{ceil}{rgb}{0.57, 0.63, 0.81}
\definecolor{antiquefuchsia}{rgb}{0.57, 0.36, 0.51}
\definecolor{bronze}{rgb}{0.8, 0.5, 0.2}
\definecolor{carrotorange}{rgb}{0.93, 0.57, 0.13}
\definecolor{coolgrey}{rgb}{0.55, 0.57, 0.67}
\definecolor{corn}{rgb}{0.98, 0.93, 0.36}
\definecolor{frenchbeige}{rgb}{0.65, 0.48, 0.36}
\definecolor{dandelion}{rgb}{0.94, 0.88, 0.19}
\definecolor{cadet}{rgb}{0.33, 0.41, 0.47}

\renewcommand{\triangleq}{\mathbin{\setstackgap{S}{0pt}\stackMath\Shortstack{\smalltriangleup\\ =}}}
\usepackage[usestackEOL]{stackengine} 
\def\BibTeX{{\rm B\kern-.05em{\sc i\kern-.025em b}\kern-.08em
    T\kern-.1667em\lower.7ex\hbox{E}\kern-.125emX}}

\begin{document}
\title{When Near Becomes Far: From Rayleigh to Optimal Near-Field and Far-Field Boundaries\\
\thanks{This work is supported in part by Digital Futures. G.Fodor was also supported by the Swedish Strategic Research (SSF) grant for the FUS21-0004 SAICOM project and the 6G-Multiband Wireless and Optical Signaling for Integrated Communications, Sensing and Localization (6G-MUSICAL) EU Horizon 2023 project, funded by the EU, Project ID: 101139176.}
}

\author{
    \IEEEauthorblockN{Sajad Daei \IEEEauthorrefmark{1},  \, Gabor Fodor \IEEEauthorrefmark{1}\IEEEauthorrefmark{2}, \, Mikael Skoglund \IEEEauthorrefmark{1}} 
    
    \IEEEauthorblockA{\IEEEauthorrefmark{1} School of Electrical Engineering $\&$ Computer Science, KTH Royal Institute of Technology, Sweden}  
    \IEEEauthorblockA{\IEEEauthorrefmark{2} Ericsson Research, Stockholm, Sweden} 
    \IEEEauthorblockA{sajado@kth.se,\, gaborf@kth.se\,, skoglund@kth.se}
}

\maketitle


\begin{abstract}
The transition toward 6G is pushing wireless communication into a regime where the classical plane-wave assumption no longer holds. Millimeter-wave and sub-THz frequencies shrink wavelengths to millimeters, while meter-scale arrays featuring hundreds of antenna elements dramatically enlarge the aperture. Together, these trends collapse the classical Rayleigh far-field boundary from kilometers to mere single-digit meters. Consequently, most practical 6G indoor, vehicular, and industrial deployments will inherently operate within the radiating near-field, where reliance on the plane-wave approximation leads to severe array-gain losses, degraded localization accuracy, and excessive pilot overhead.
This paper re-examines the fundamental question: \emph{``Where does the far-field truly begin?''} Rather than adopting purely geometric definitions, we introduce an application-oriented approach based on user-defined error budgets and a rigorous Fresnel-zone analysis that fully accounts for both amplitude and phase curvature. We propose three practical mismatch metrics: worst-case element mismatch, worst-case normalized mean square error, and spectral efficiency loss. For each metric, we derive a provably optimal transition distance--the minimal range beyond which mismatch permanently remains below a given tolerance--and provide closed-form solutions. Extensive numerical evaluations across diverse frequencies and antenna-array dimensions show that our proposed thresholds can exceed the Rayleigh distance by more than an order of magnitude. By transforming the near-field from a design nuisance into a precise, quantifiable tool, our results provide a clear roadmap for enabling reliable and resource-efficient near-field communications and sensing in emerging 6G systems.

\end{abstract}


\makeatletter{\renewcommand*{\@makefnmark}{}
\footnotetext{This work is supported in part by Digital Futures. G.Fodor is also supported by the Swedish Strategic Research (SSF) grant for the FUS21-0004 SAICOM project and the 6G-Multiband Wireless and Optical Signaling for Integrated Communications, Sensing and Localization (6G-MUSICAL) EU Horizon 2023 project, funded by the EU, Project ID: 101139176.
S.Daei is supported by Digital Futures Project STACEY. }
\makeatother}

\section{Introduction}\label{sec:intro}


\noindent\textbf{From rule of thumb to rule breaker.}%
\;Diffraction theory's most enduring yardstick dates back to 1903, when Lord Rayleigh demonstrated that the Kirchhoff integral for a circular aperture simplifies to a plane-wave (Fraunhofer) form whenever the Fresnel number falls below one-half. The corresponding range,
\begin{equation}
  R_{\mathrm{Ray}} \triangleq \frac{2D^{2}}{\lambda},
\end{equation}
with aperture \(D\) and wavelength \(\lambda\), became the canonical \emph{Rayleigh distance} and has been cited verbatim in virtually every antenna text and measurement standard to date \cite{rayleigh1896xv,born2013principles,stutzman2012antenna}.
For the high frequency (HF) and very high frequency (VHF) systems that dominated the past century--with
array apertures of tens of centimeters and wavelengths of
several meters--this distance comfortably exceeded practical
link lengths, justifying widespread reliance on the plane-wave
assumption \cite{balanis2016antenna,johnson1984antenna,hurd1980ieee,rappaport2017overview}.

However, as wireless systems evolve towards sixth-generation (6G) networks, the classical geometry underpinning Rayleigh’s criterion has drastically changed. Millimeter-wave (mmWave) and sub-terahertz (THz) frequencies compress wavelengths down to mere millimeters, while modern antenna arrays--spanning hundreds of elements--grow to meter-scale dimensions. Consequently, the Rayleigh distance shrinks from kilometers to mere meters, placing typical indoor, industrial, and vehicular deployments firmly within what classical antenna theory still labels the \emph{radiating near-field}\cite{rappaport2017overview,alkhateeb2014channel,liu2017millimeter}.

\noindent\textbf{Why a ``plane-wave'' mindset is no longer sufficient.}  
Rayleigh’s classical criterion was originally designed to limit only phase errors (to approximately \(22.5^\circ\)), leaving several key phenomena unaddressed that become significant at 6G frequencies \cite{cui2022channel}. At millimeter-wave and sub-THz frequencies, amplitude curvature emerges prominently, producing power variations of up to \(6\,\text{dB}\) across meter-scale arrays and thereby undermining beamforming techniques premised on uniform amplitude distributions. Furthermore, the mismatch between spherical and planar wavefronts does not follow a simple monotonic decline; instead, it exhibits counterintuitive oscillations before ultimately diminishing--a nuance that conventional small-angle approximations fail to capture. Finally, different error metrics, such as worst-case element mismatch, normalized mean square error (NMSE), and spectral efficiency (SE) loss--vary sharply in their criteria for acceptable mismatch, directly impacting localization accuracy, beamforming efficiency, and overall throughput performance.

\noindent\textbf{Near-field is an opportunity, not an inconvenience.}
Rather than a nuisance, the radiating near-field region opens three distinctive avenues for 6G innovation \cite{alkhateeb2014channel,shahmansoori2017position,cui2022channel,daei2025near}.
First, the pronounced spherical curvature enables \emph{three-dimensional beam focusing}, delivering centimeter-level indoor-positioning accuracy without resorting to ultra-wide bandwidths.
Second, because each user possesses a unique amplitude and phase signature that depends on range as well as azimuth, a base station (BS) can perform \emph{depth-resolved multiplexing}: users sharing the same angle but located at different distances become separable, boosting line-of-sight throughput well beyond what planar beams permit.
Third, the deterministic amplitude profile embedded in every packet can be repurposed for \emph{zero-overhead sensing}; the same waveform that conveys data simultaneously serves as a monostatic radar snapshot, with no extra pilots, dwell time, or equivalent isotropically radiated power (EIRP)\cite{daei2024timely,razavikia2023off}. Translating these capabilities into deployable features demands near–to-far-field boundaries expressed in engineering units--decibels, bits/s/Hz, or centimeters of localization error, rather than a purely geometric threshold.

\subsection{Contributions}\label{subsec:ctrb}
This paper introduces three application-driven mismatch metrics and corresponding optimal boundaries that comprehensively redefine the near-to-far field transition for modern wireless systems:

\begin{enumerate}
    \item \textbf{Worst-case Element-wise Mismatch (\(\ell_{\infty}\))}: Targets hardware-constrained systems by highlighting the maximum deviation at any single antenna element, critical for RF-chain calibration and front-end linearity design.
    
    \item \textbf{Worst-case Normalized \(\ell_{2}\) Mismatch}: Essential for channel estimation accuracy, positioning performance, and multi-user precoding; this metric quantifies total array-gain degradation due to model mismatch.
    
    \item \textbf{Spectral Efficiency Loss}: Directly measures the throughput penalty (bits/s/Hz) incurred by using far-field beamforming strategies in near-field environments, guiding link-budgeting, resource allocation, and integrated sensing and communication (ISAC) trade-offs\cite{daei2024timely,daei2025one}.
\end{enumerate}

For each metric, we rigorously derive an optimal threshold distance--the minimal range beyond which the specified mismatch remains permanently below a designer-defined tolerance. We also propose three angle-independent analytic thresholds: the Exact-Phase Fresnel (EPF), Small-Phase Fresnel (SPF) and Strict SPF (SSPF) boundaries--with the SSPF boundary given explicitly as: $  R_{\rm SSPF}\triangleq \sqrt{\tfrac{k_{\lambda}D^2+D}{2\delta_{\infty}}}
$,
where \(k_{\lambda}\triangleq 2\pi/\lambda\) is the wavenumber, and \(\delta_{\infty}\) is the mismatch tolerance dictated by the target application.
Extensive numerical evaluations spanning carrier frequencies from \(0.1\,\text{GHz}\) up to \(300\,\text{GHz}\) and array sizes up to \(64\) antenna elements confirm that our proposed application-specific boundaries substantially surpass the classical Rayleigh distance--often by more than an order of magnitude at mmWave and sub-THz bands. By shifting the near-field from an ambiguous, uncontrolled regime to a precisely quantified and controllable design parameter, these findings provide a rigorous and practical engineering roadmap, providing robust and resource-efficient wireless communications and sensing in future 6G deployments.

\begin{figure}[!t]
  \centering
  \begin{subfigure}[b]{0.19\textwidth}
    \includegraphics[scale=.12,trim={0cm 0 0 0}]{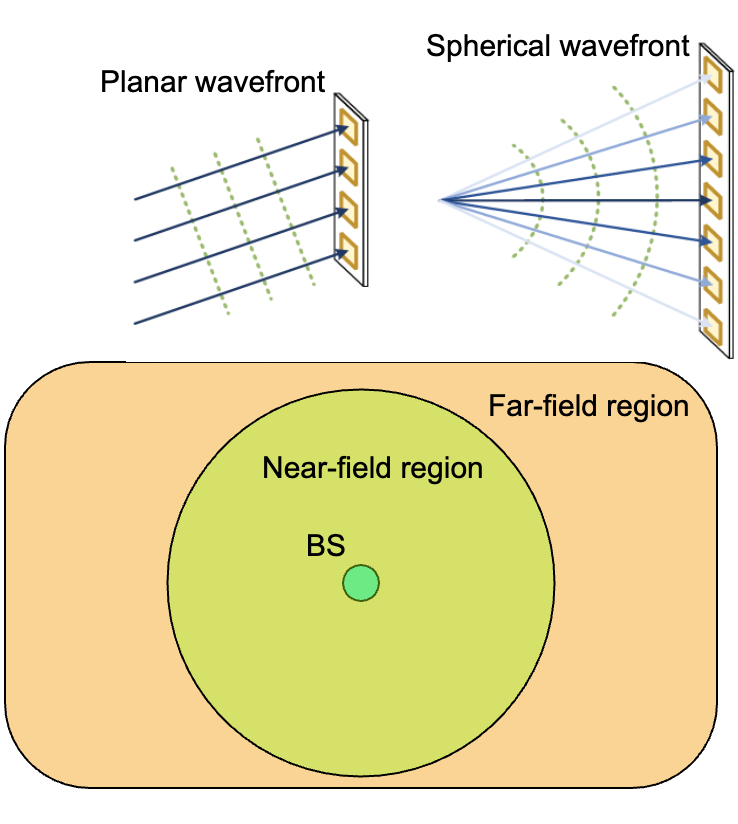}
    \caption{}
      \label{fig:schematic}
  \end{subfigure}%
  \hfill
  \begin{subfigure}[b]{0.19\textwidth}
    \includegraphics[scale=.18,trim={12cm 0 0 0}]{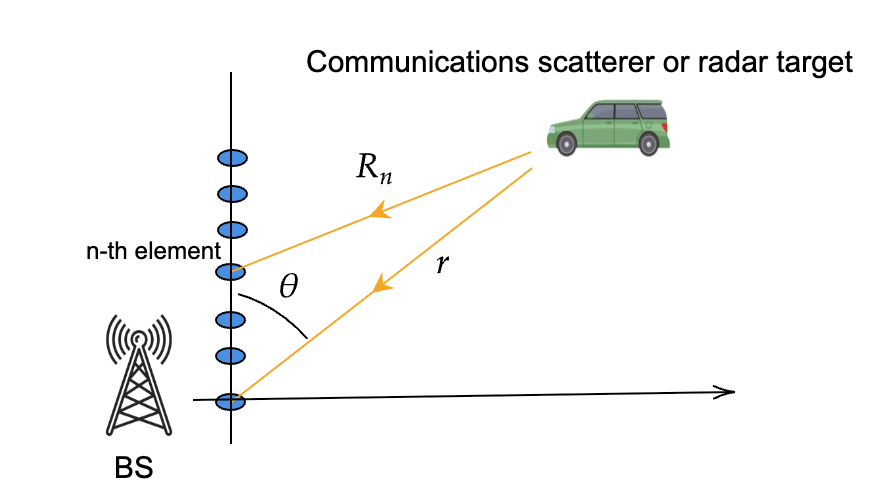}
    \caption{}
    \label{fig:model}
  \end{subfigure}%

  \label{fig:allBounds}
  \caption{(a) Top: Planar wavefront in far-field versus spherical wavefront in near-field. Bottom: This figure shows the near-field region around a BS which is known as Rayleigh distance. (b) System model under consideration, featuring a BS equipped with a multi-antenna array and a single remote node. From a communications perspective, this single-antenna node actively transmits uplink data signals toward the BS. From a sensing (radar) perspective, the same node can alternatively act as a passive radar target, reflecting an incident probing signal back toward the BS.}
\end{figure}

\textbf{Roadmap.}  
The remainder of this paper is organized as follows. Section~\ref{sec:system-model} presents the system model. Section~\ref{sec:proposed_method} introduces the proposed mismatch metrics along with their corresponding optimal transition distances. Section \ref{sec:app} explains the applications of the proposed mismatch metrics. In Section~\ref{sec:numerical_results}, extensive numerical experiments validate the performance of our application-driven boundaries and compare them against the classical Rayleigh distance. Finally, Section~\ref{sec:conclusion} concludes the paper and highlights promising future research directions toward practical 6G wireless deployments.

\textbf{Notation}.
Vectors are denoted by small bold letters and matrices are denoted by capital bold letters. The growth of a function \(f(r)\) relative to another function \(g(r)\) is expressed as $f(r) = \mathcal{O}\bigl(g(r)\bigr),$
which means that $\exists C>0,r_0:  |f(r)|\le C\,|g(r)|~\forall r\ge r_0$. The $n$-th element of a vector $\mx{a}\in\mathbb{C}^{N\times 1}$ that depends on the parameter $\theta$ is shown by $a(\theta)[n]$. The $\ell_{\infty}$ and $\ell_2$ norms of a vector $\mx{x}\in\mathbb{C}^{N\times 1}$ are defined respectively as $\|\mx{x}\|_{\infty}\triangleq \max_n |x[n]|$ and $\|\mx{x}\|_{2}\triangleq \sqrt{\sum_{n=1}^N |x[n]|^2}$. $j\triangleq\sqrt{-1}$ denotes the imaginary unit.

\section{System Model}

\label{sec:system-model}
We consider a communication user that transmits a signal toward a BS equipped with $N_r$ number of antennas. The array of antennas located on the y-axis in Figure \ref{fig:model} forms a uniform linear array (ULA) of $N_r$ isotropic elements, spanning an aperture $D=(N_r-1)d$ with inter-element spacing $d=\lambda/2$. The user is located at polar angle $\theta$ and range $r$ measured from the array first element located at the origin $(0,0)$.  The channel is considered to be line-of-sight (LoS) from the user to the BS. From a radar perspective, the user could instead be viewed as a radar target reflecting signals emitted by a radar transmitter toward the BS. The carrier frequency is given by $f_c=\tfrac{c}{\lambda}$ where $c$ is the speed of light.

\subsection{Near-field Steering Vector}
The $n$-th element of the near-field steering vector is given by\cite{cui2022channel,zhang2023near}:
\begin{align}\label{eq:near_steering}
    a_{\mathrm{NF}}(\theta,r)[n]\;\triangleq\;
 {\rm e}^{-j\,k_{\lambda} R_{n}},
 \quad
 n=0,\dots,N_r-1,
\end{align}
with exact distance
$R_{n}=\sqrt{\,r^2 + (n d)^2 - 2\,r n d\cos(\theta)}$.
Stacking all $N_r$ samples yields the \emph{near-field steering vector}
$\mx{a}_{\mathrm{NF}}(\theta,r)\in\mathbb{C}^{N_r\times 1}$.

In the far-field ($r\!\to\!\infty$), the phase becomes linear, and the near-field steering vector converts to the conventional far-field which is given by:
\begin{align}\label{eq:far_steering}
a_{\mathrm{FF}}(\theta,r)[n]
={\rm e}^{-jk_{\lambda}(r-n d\cos\theta)}~ \forall n=0,..., N_r-1   
\end{align}

The uplink baseband channel from the user located in the near-field and far-field region to the BS then becomes of the following forms\cite[Eq. 1]{zhang2023near}:
\begin{align}
  &{h}_{\rm NF}[n]=
      \sqrt{g_Tg_R[n]}\tfrac{\lambda}{4\pi R_n}\,{a}_{\mathrm{NF}}(\theta,r)[n]\\
     &{h}_{\rm FF}[n]=
      \sqrt{g_T g_R}\tfrac{\lambda}{4\pi r}\,{a}_{\mathrm{FF}}(\theta,r)[n] 
  \;\;n=0,..., N_r-1,
  \label{eq:channel}
\end{align}
where $g_T$ and $g_R[n]$ are the transmit and receive antenna gain. $r\triangleq R_{0}$ is the distance of the user/target from the first antenna located at the origin.
For ease of analysis in this paper, we assume that $g_T=g_R[n]=1$. This means that the elements of the array are isotropic.

\section{Proposed method}\label{sec:proposed_method}
In this section, we introduce three metrics to evaluate the mismatch between near-field and far-field and propose the optimal thresholds-- the distances beyond which the mismatch metric remains below a given threshold.

\subsection{Worst-Case Element-Wise Mismatch}

The worst-case element-wise mismatch is defined as below:
\begin{align}\label{eq:worst-case_mismatch}
 &E_{\ell_{\infty}}(r)\triangleq \nonumber\\
 &\max_{\theta\in(0,2\pi)}\max_{0\le n\le N_r-1} \bigl|\tfrac{1}{R_n} {\rm e}^{-j k_{\lambda} R_n}-\tfrac{1}{r} {\rm e}^{-j k_{\lambda}(r-  n d \cos(\theta))} \bigr|\nonumber\\
&\triangleq \max_{\theta\in(0,2\pi)} E_{\ell_{\infty}}(r,\theta).    
\end{align}
The metric $ E_{\ell_{\infty}}(r)$ is the largest $\ell_{\infty}$ mismatch seen by any single antenna.

\begin{prop}\label{prop.optimal_boundary}
Let \(E_{\ell_{\infty}}(r)\) denote the worst–case element-wise mismatch
defined in \eqref{eq:worst-case_mismatch},
and fix an application-driven tolerance
\(\delta_{\infty}>0\).

{(i) Optimal distance.}
The smallest range at which the metric \emph{never} again exceeds the
tolerance is
\begin{equation}
  R_{\mathrm{OPT},\ell_{\infty}}
  \triangleq 
  \inf\!\bigl\{r\ge0:\;\sup_{r'\ge r}E_{\ell_{\infty}}(r')<\delta_{\infty}\bigr\}.
  \label{eq:Ropt_infty}
\end{equation}

{(ii) EPF distance.}
Using Fresnel approximation and an upper-bound of $E_{\ell_{\infty}}(\cdot)$ leads to the EPF distance given by
\begin{equation}
R_{\mathrm{EPF}}
\triangleq
\inf\Bigl\{r\ge 0:\;
\sup_{r'\ge r}
\Bigl[\tfrac{D^{2}}{2r'^{3}}
      +\tfrac{2}{r'}\bigl|\sin(\tfrac{k_{\lambda}D^{2}}{4r'})\bigr|\Bigr]
      \le\delta_{\infty}
\Bigr\}.
\label{eq:EPF}
\end{equation}

{(iii)  SPF distance.}
Using the small angle approximation \(\sin x\le x\) yields the following closed-form expression:
\begin{align}
&R_{\mathrm{SPF}} \triangleq 2\sqrt{\tfrac{k_{\lambda} D^2}{6\delta_{\infty}}}\cos(\tfrac{1}{3}\arccos(\tfrac{3}{2k_{\lambda}}\sqrt{\tfrac{6\delta_{\infty}}{k_{\lambda} D^2}})).
\label{eq:SPF}
\end{align}

{(iv) SSPF distance.}
Imposing the physical constraint \(r\ge D\) provides the conservative
yet simplified bound
\begin{equation}
  R_{\mathrm{SSPF}}
  \triangleq
  \sqrt{\tfrac{k_{\lambda}D^{2}+D}{2\delta_{\infty}}}.
  \label{eq:SSPF}
\end{equation}
These thresholds allow designers to balance analytic simplicity against tightness
according to system requirements.
\end{prop}
Proof. See the appendix.
\begin{rem}(Proposed boundaries)
The \emph{optimal} transition distance introduced in this work is obtained without resorting to any high-frequency or small-angle approximations: it is the unique largest range beyond which the mismatch metric never again exceeds the tolerance.  Existence is guaranteed because \(E_{\ell_{\infty}}(r)\) is continuous in \(r\) and \(\lim_{r\to\infty}E_{\ell_{\infty}}(r)=0\).
The EPF radius in \eqref{eq:EPF} is a tractable approximation to this optimal threshold.  It follows from retaining the full Fresnel phase term and solving the resulting nonlinear equation.
The SPF bound in \eqref{eq:SPF} applies the small-angle inequality \(\sin x\le x\), yielding a fully closed-form expression.  By additionally enforcing the physical constraint \(r\ge D\), we obtain the SSPF radius in \eqref{eq:SSPF}, which is a conservative yet still analytic bound. 
\end{rem}
\vspace{.2cm}
\subsection{Worst-Case Normalized $\ell_2$ Mismatch}

    The normalized $\ell_2$ mismatch metric is defined as:
    
\begin{align*}
     &E_{\ell_{2}}(r)\triangleq \max_{\theta\in(0,2\pi)} \tfrac{\sqrt{\sum_{n=0}^{N_r-1} \bigl|\tfrac{1}{R_n} {\rm e}^{-j k_{\lambda} (R_n-r)}-\tfrac{1}{r} {\rm e}^{j k_{\lambda} n d \cos(\theta)} \bigr|^2}}{\sqrt{\sum_{n=0}^{N_r-1} \bigl|\tfrac{1}{R_n} {\rm e}^{-j k_{\lambda} R_n}\bigr|^2}}\nonumber\\
     &\triangleq \max_{\theta\in(0,2\pi)}  E_{\ell_{2}}(r,\theta)
\end{align*}
    and it 
captures the total array‐gain difference between far-field and near-field. We also call this NMSE mismatch. Similar to Proposition \ref{prop.optimal_boundary}, we can define the optimal distance as the minimum range after which this metric does not exceed a given threshold $\delta_2$ as follows: 
\begin{align}\label{eq:R_opt,ell2}
   R_{{\rm OPT},\ell_2}\triangleq 
  \inf\!\bigl\{r\ge0:\;\sup_{r'\ge r}E_{\ell_{2}}(r')<\delta_{2}\bigr\},
\end{align}
where $\delta_2$ is a dimensionless NMSE tolerance. 
\begin{remark}[Closed-form vs.\ practical evaluation of $R_{\mathrm{OPT},\ell_2}$]
Unlike the element-wise threshold $R_{\mathrm{OPT},\ell_\infty}$, the normalized $\ell_2$ threshold 
$R_{\mathrm{OPT},\ell_2}$ does not admit a simple Rayleigh-style closed form, as it involves a maximization 
over oscillatory Fresnel-phase terms. In practice, however, it can be determined efficiently through a lightweight procedure: 
(i) sweep the range $r$ on a grid, (ii) maximize the mismatch over $\theta$, and 
(iii) identify the smallest $r$ beyond which the ``never-again exceeds'' condition in (11) holds. 
This is essentially a one-dimensional search in~$r$ with an angular maximization, which is computationally inexpensive. 
In addition, conservative \emph{closed-form sufficient bounds} can be obtained by relating $E_{\ell_2}$ to $E_{\ell_\infty}$ via norm inequalities, 
providing safe radius estimates. Refining these into tighter closed-form approximations remains an interesting direction for future work.
\end{remark}
 
\begin{rem}(Bias floor versus the $\ell_2$ mismatch)\label{rem.NMSE_bias}

Consider a single-antenna user, transmitting the zero-mean pilot signal $\mx{x}\in\mathbb{C}^{L\times 1}$ to the BS ($\mathbb{E}[|x[l]|^2]=P_p$). The BS collects the measurements as $\mx{Y}=\mx{h}_{\rm NF} \mx{x}^T+\mx{W}\in\mathbb{C}^{N_r\times L}$ where $\mx{W}$ is additive noise matrix. Now, suppose the BS wrongly believes the user lies in the far-field region and solves the constrained least square problem $\widehat{\mx{h}}=\mathop{\arg\min}_{\mx{h}\in{\rm span}(\mx{a}_{\rm FF})} \|\mx{Y}-\mx{h}\mx{x}^H\|_F^2$ to estimate the channel which leads to $\widehat{\mx{h}}=\tfrac{\mx{h}^H_{\rm FF} \mx{h}_{\rm NF}}{\|\mx{h}_{\rm FF}\|_2^2} \mx{h}_{\rm FF}+\mx{n}$ where $\mx{n}$ is a zero-mean Gaussian random variable with $\mathbb{E}[\mx{n}\mx{n}^H]=\tfrac{1}{L\rho_P}\tfrac{\mx{h}_{\rm FF}\mx{h}_{\rm FF}^{\mathsf{H}}}{\|\mx{h}_{\rm FF}\|_2^2}$ and $\rho_{\rm p}\triangleq\tfrac{P_p}{\sigma_p^2}$ is the pilot SNR. A lower-bound for the NMSE of the channel estimate can be expressed as:
\begin{align}\label{eq:nmse_me}
{\rm NMSE}\triangleq\mathds{E}_{\mx{W}} [\tfrac{\|\mx{h}_{\rm NF}-\widehat{\mx{h}}\|_2^2}{\|\mx{h}_{\rm NF}\|_2^2}]\ge \underbrace{{(1-\eta)}}_{\triangleq {\rm NMSE}_{\rm bias}}+    \tfrac{1}{L \rho_{\rm p} \|\mx{h}_{\rm NF}\|_2^2}
\end{align}
where $  \eta(r,\theta)
  \triangleq \tfrac{\bigl|\mx{h}_{\mathrm{NF}}^{H}\,\mx{h}_{\mathrm{FF}}\bigr|^{2}}
         {\|\mx{h}_{\mathrm{NF}}\|_{2}^{2}\,\|\mx{h}_{\mathrm{FF}}\|_{2}^{2}}
  \;\in[0,1]$
is refereed to as the \emph{array‐gain efficiency}. When amplitude curvature is negligible, \(E_{\ell_{2}}^{2}\!\approx\!1-\eta\), and the bias term can be approximated by:
\begin{align}\label{eq:NMSE_bias}
\mathrm{NMSE}_{\text{bias}}
     \;\approx
     1-\Bigl(1-\tfrac{E_{\ell_{2}}^{2}}{2}\Bigr)^{2} \approx \;E_{\ell_{2}}^{2}+{\cal O}\!\bigl(E_{\ell_{2}}^{4}\bigr).
\end{align}  
\end{rem}
\subsection{Spectral Efficiency Mismatch}
 Consider a single-antenna user transmitting data symbol \(s\) (with \(\mathbb E[|s|^2]=P_{\rm d}\)) to an \(N_r\)-element BS and define $G\triangleq \|\mx{h}_{\rm NF}\|_2^2$.  The received vector at the BS is
\begin{equation}
  \mx{y} \;=\;\mx{h}_{\mathrm{NF}}(r,\theta)\,s \;+\;\mx{w},\quad
  \mx{w}\sim \mathcal{CN}(\mathbf0,\sigma_{\rm d}^2\mx{I})\in\mathbb{C}^{N_r\times 1},
\end{equation}
where $\mx{w}$ is noise vector at data time slot, \(\mx{h}_{\mathrm{NF}}\in\mathbb{C}^{N_r\times 1}\) is the true near-field channel steering vector as in \eqref{eq:worst-case_mismatch}, and \(\rho_{\rm d}=P_d/\sigma_{\rm d}^2\) denotes the data SNR. With perfect CSI, the BS applies the optimal matched filter $\mx{v}={\mx{h}}_{\rm NF}$, yielding post-combiner SNR ${\rm SNR}_{\rm opt}=\rho G$ and the following spectral efficiency \cite{shannon1948mathematical}:
\begin{equation}
  {\rm SE}_{\mathrm{opt}}
  \triangleq \log_{2}\bigl(1 +{\rm SNR}_{\rm opt}\bigr).
\end{equation}
If instead the BS (mistakenly) uses the planar subspace ${\rm span}(\mx{a}_{\rm FF})$ in the constrained least square problem for channel estimation and employs it as its combiner, the resulting SNR becomes
\begin{align}
   \mathrm{SNR}_{\mathrm{mis}}
  = \tfrac{\eta^2 G^2 \rho_{\rm d}}{\tfrac{G \eta\rho_{\rm d}}{L \rho_{\rm p}}+\eta G+\tfrac{1}{L\rho_{\rm p}}}. 
\end{align}
The corresponding spectral efficiency under mismatch reads as ${\rm SE}_{\mathrm{mis}}
  = \log_{2}\bigl(1 + \mathrm{SNR}_{\mathrm{mis}}\bigr)$.
The \emph{SE mismatch} or \emph{SE-loss} is defined as the difference $\Delta {\rm SE}(r,\theta)
  \triangleq {\rm SE}_{\mathrm{opt}} - {\rm SE}_{\mathrm{mis}}$. This metric quantifies, in bits/s/Hz, the spectral efficiency penalty incurred by planar-wave channel estimation and beamforming design in true near‐field conditions.
To ensure a guaranteed SE performance over all look angles, we define the \emph{worst-case SE loss}
\begin{align}\label{eq:SEloss}
  E_{\rm SE}(r)
  \triangleq\max_{\theta\in(0,2\pi)} \Delta {\rm SE}(r,\theta),
\end{align}
and select the \emph{SE-safe distance} \(R_{{\rm OPT},{\rm SE}}\) as the smallest \(r\) beyond which the SE mismatch does not exceed an SE tolerance $\delta_{\rm SE}$ i.e.,
\begin{align}\label{eq:R_opt.se}
   R_{{\rm OPT},{\rm SE}}\triangleq\inf\{r:  \sup_{r'\ge r}E_{\rm SE}(r')\le\delta_{\rm SE}\}.
\end{align}
As with $R_{\mathrm{OPT},\ell_2}$, the SE-based threshold $R_{\mathrm{OPT},{\rm SE}}$ cannot be expressed 
in a simple Rayleigh-style closed form, but it can be efficiently computed via the same one-dimensional 
search procedure, while conservative closed-form bounds remain a topic for future investigation.
\begin{rem}[Range dependence of $E_{\ell_\infty}$, $E_{\ell_2}$, and SE loss]
For any fixed angle~$\theta$, both the element-wise mismatch $E_{\ell_\infty}(r,\theta)$ and the normalized $\ell_2$ mismatch $E_{\ell_2}(r,\theta)$ inherit an \emph{oscillatory} dependence on the range~$r$ from the Fresnel phase curvature retained in the exact-phase analysis. In particular, the phase term in \eqref{eq:strict_condition} contains $\bigl|\sin\!\bigl(\tfrac{k_{\lambda} D^2}{4r}\bigr)\bigr|$, which produces periodic re-phasing and de-phasing of the near-far mismatch as $r$ varies. As a result, $E_{\ell_\infty}(r,\theta)$ and $E_{\ell_2}(r,\theta)$ are generally \emph{non-monotonic} in~$r$ and vanish as $r \to \infty$. The same applies to the SE loss $\Delta\mathrm{SE}(r,\theta)$ in \eqref{eq:SEloss}, since $\mathrm{SE}_{\mathrm{mis}}$ is monotone in the array-gain efficiency $\eta(r,\theta)$, and $\eta$ is the squared, normalized inner product between $\mx{h}_{\mathrm{NF}}$ and $\mx{h}_{\mathrm{FF}}$.
When maximizing over~$\theta$, the second-order expansion (see \eqref{eq:phase_term}) shows that the dominant phase-curvature term scales with $\sin^2\theta$, so for practically relevant ranges where this term dominates, the worst case occurs near $\theta \approx \pm\pi/2$. At short ranges, the amplitude term in \eqref{eq:amplitude_term}, which scales with $|\cos\theta|$, may dominate and shift the maximizer toward $\theta \approx 0$ or~$\pi$. Furthermore, the maximizing angle can switch as $r$ varies, causing local increases in the worst-case curves. These behaviors motivate defining $R_{\mathrm{OPT},\ell_\infty}$, $R_{\mathrm{OPT},\ell_2}$, and $R_{\mathrm{OPT},\mathrm{SE}}$ via the ``never-again exceeds'' envelope in \eqref{eq:Ropt_infty}, \eqref{eq:R_opt,ell2}, and \eqref{eq:R_opt.se}, which is monotone by construction and robust to such oscillations.

\end{rem}
\section{Applications}\label{sec:app}
A single, fixed geometric boundary is insufficient to satisfy the heterogeneous requirements that arise across the 6G protocol stack. Accordingly, we tailor each design layer to the metric and transition distance that most directly governs its performance.

\subsection{RF Hardware: Worst-Element Metric \(E_{\ell_\infty}\)}
Guaranteeing  
\(E_{\ell_\infty}(r)\le\delta_\infty\Rightarrow r\ge R_{\mathrm{OPT},\ell_\infty}\)  
(i) keeps phase shifters and low noise amplifiers (LNAs) within their quantization and linearity
budgets \cite{molisch2017hybrid},  
(ii) sets the analog to digital converter (ADC) back-off required for spectrum emission mask (SEM) limits
\cite{han2015large},  
(iii) meets per-element EIRP and specific absorption rate (SAR) limits \cite{docomo20165g}, and  
(iv) isolates drifting RF chains through real-time monitoring.

\subsection{Baseband: Normalized NMSE Metric \(E_{\ell_2}\)}
The energy error \(E_{\ell_2}(r)\) bounds the array-gain loss.
Keeping \(E_{\ell_2}\le\delta_{2}\Rightarrow r\ge R_{\mathrm{OPT},\ell_2}\)  
(i) fixes link-budget margins and modulation coding scheme (MCS) tables (3 dB loss \(\approx\) 1 bit/s/Hz at high SNR \cite{rappaport2017overview}),  
(ii) limits the NMSE floor of LS/MMSE estimators to \(E_{\ell_2}^{2}\)(see Remark \ref{rem.NMSE_bias} and Equation \ref{eq:nmse_me}) \cite{lu2014overview},  
(iii) preserves Fisher-information for cm-level positioning
\cite{shahmansoori2017position}, and  
(iv) guides pilot reuse and user grouping in massive multiple-input multiple-output (MIMO) scheduling.

\subsection{Network: Spectral-Efficiency Loss \(E_{\text{SE}}\)}
The worst-case SE penalty
\(E_{\text{SE}}(r)\)
defines the SE-safe distance \(R_{\mathrm{OPT,SE}}\).  
Constraining \(E_{\text{SE}}\)  
(i) preserves the target block-error rate (BLER) in adaptive MCS \cite{love2005limited};,  
(ii) translates directly into the dB margin required to support cell-edge users \cite{alkhateeb2014channel}  
(iii) prompts the scheduler to assign more robust resource blocks--or to insert extra pilot symbols--whenever a link exceeds the allocated budget \cite{liu2017millimeter}; and
(iv) within ISAC operation, triggers a temporary switch to spherical combining during radar snapshots, thereby retaining millimeter-level sensing accuracy while maintaining full data throughput.

In summary, the triplet
\(\{R_{\mathrm{OPT},\ell_\infty},
  R_{\mathrm{OPT},\ell_2},
  R_{\mathrm{OPT,SE}}\}\)
acts as a metric-aligned “traffic light,” enabling hardware engineers, digital signal processing (
DSP) architects, and network planners to adopt the correct far-field
assumption only when their own performance budget is met.

\section{Numerical Results}\label{sec:numerical_results}

This section quantifies how closely the proposed analytical thresholds--EPF, SPF, and their strict variant SSPF--track the \emph{optimal} transition
distance obtained via precise numerical optimization, and how all of them compare
with the classical Rayleigh rule.  Unless stated otherwise, the inter-element
spacing is \(d=\lambda/2\) and the tolerance levels are
\(\delta_{\infty}=10^{-3}\,{\rm m}^{-1}\),
\(\delta_{2}=10^{-3}\),
and \(\delta_{\rm SE}=10^{-3}\) bits/s/Hz.

\begin{figure*}[!t]
  \centering
  \begin{subfigure}[b]{0.32\textwidth}
    \includegraphics[width=\linewidth]{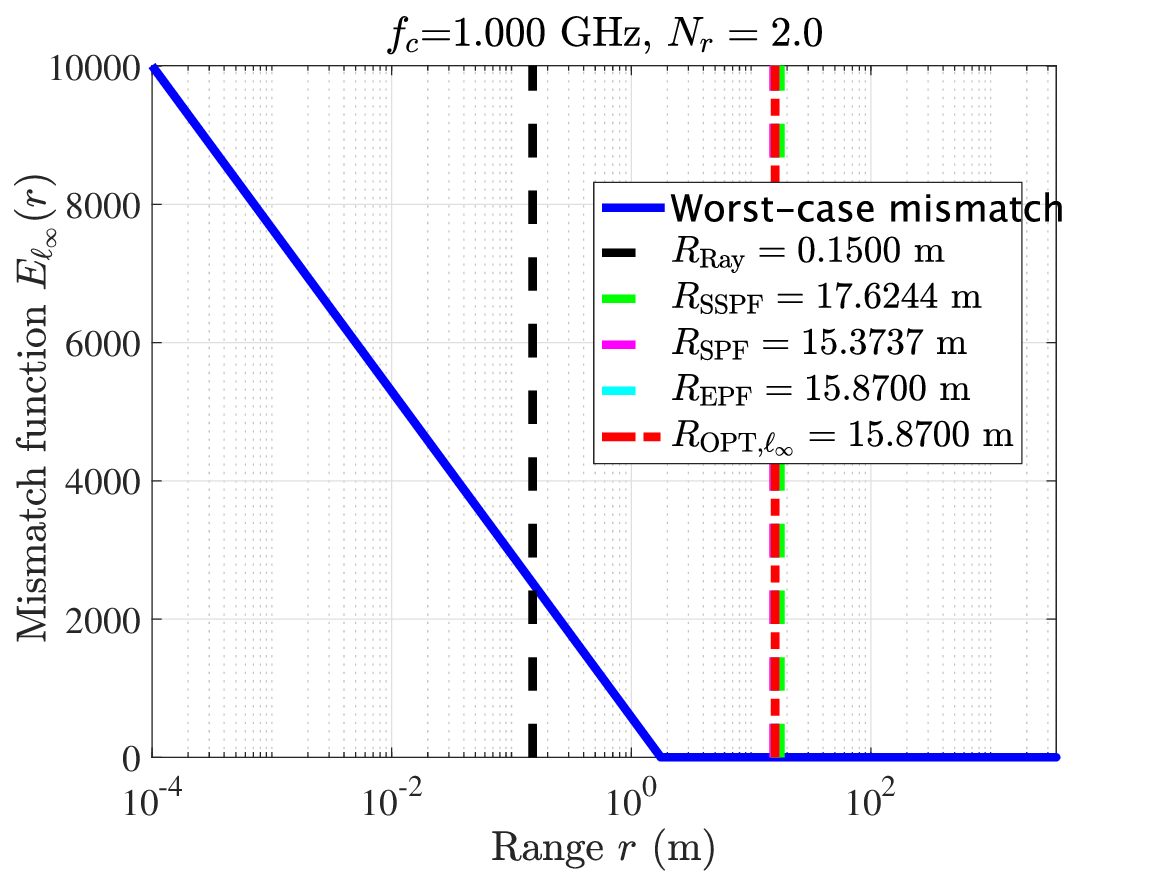}
    \caption{}
    \label{fig:l_infa}
  \end{subfigure}%
  \hfill
  \begin{subfigure}[b]{0.32\textwidth}
    \includegraphics[width=\linewidth]{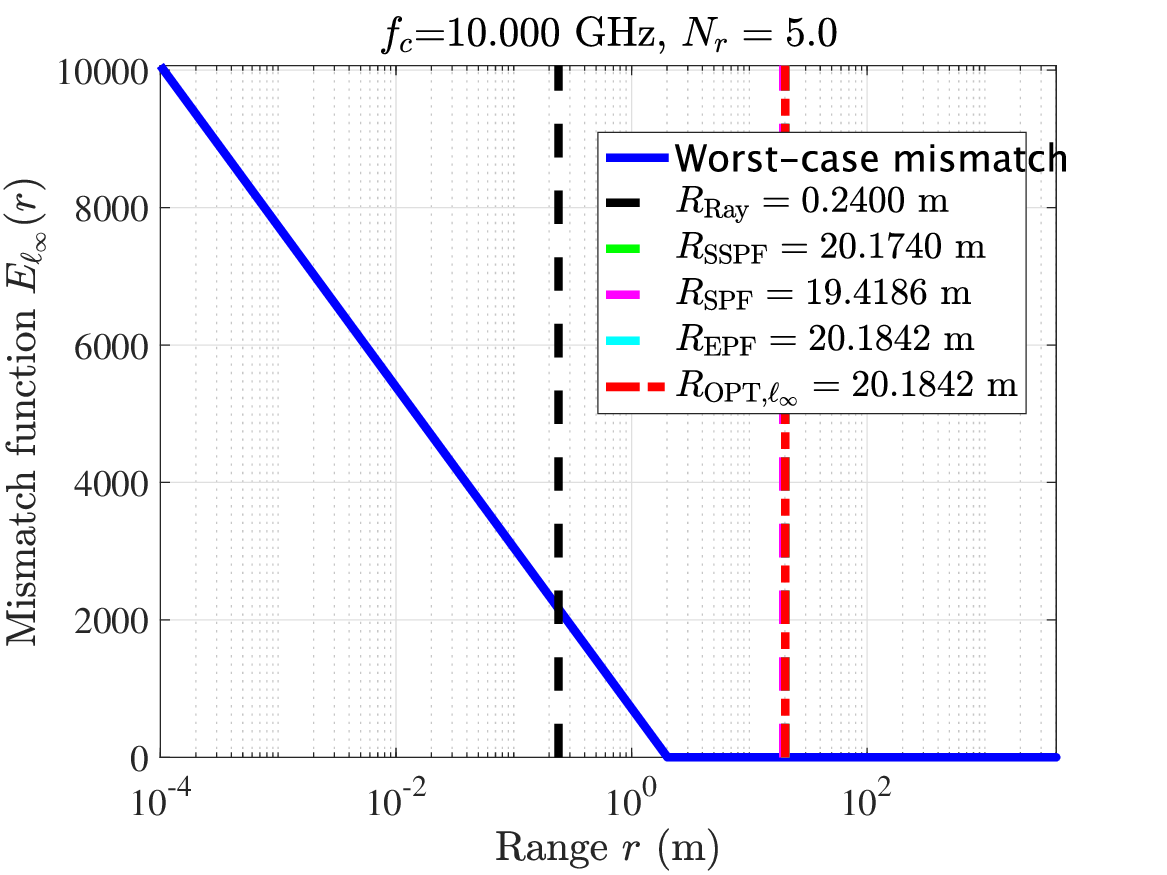}
    \caption{}
    \label{fig:l_infb}
  \end{subfigure}%
  \hfill
  \begin{subfigure}[b]{0.32\textwidth}
    \includegraphics[width=\linewidth]{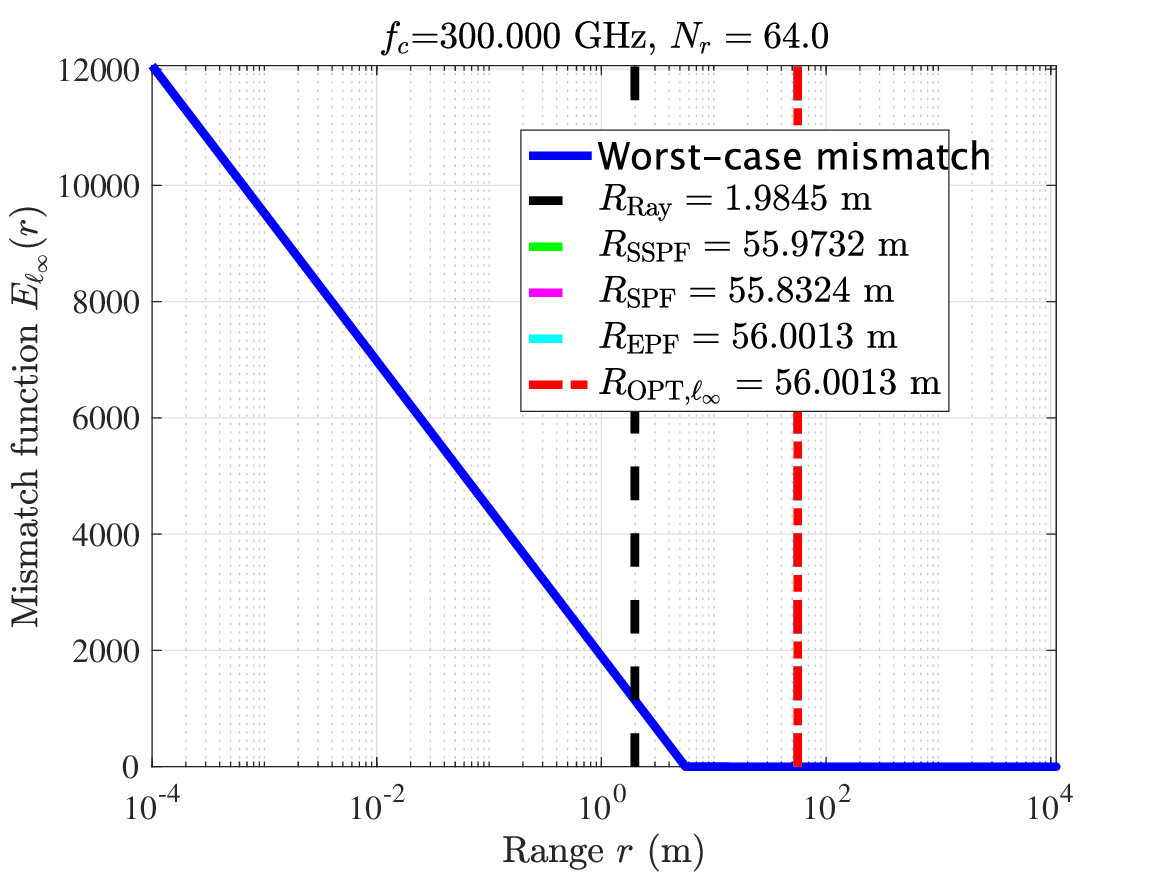}
    \caption{}
    \label{fig:l_infc}
  \end{subfigure}
  \hfill
  \begin{subfigure}[b]{0.32\textwidth}
    \includegraphics[width=\linewidth]{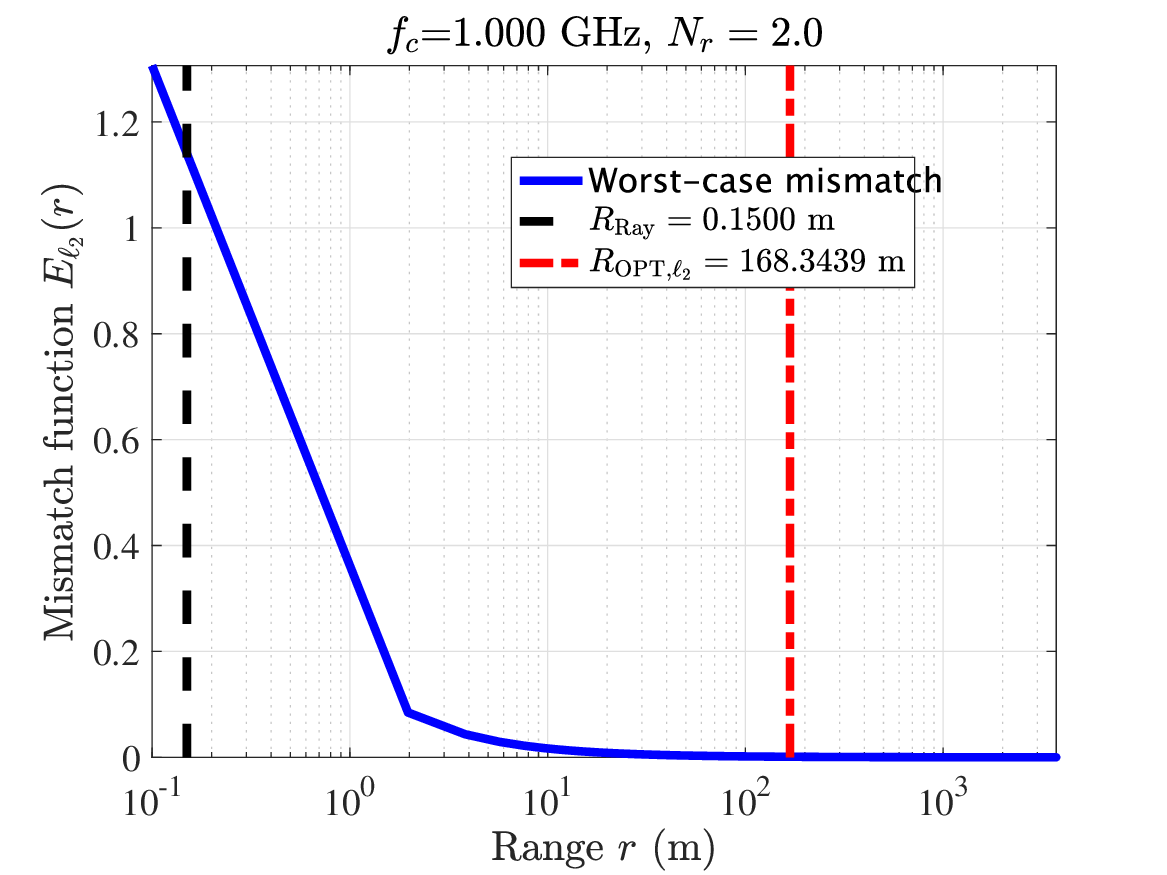}
    \caption{}
    \label{fig:l2_a}
  \end{subfigure}
 \hfill
  \begin{subfigure}[b]{0.32\textwidth}
    \includegraphics[width=\linewidth]{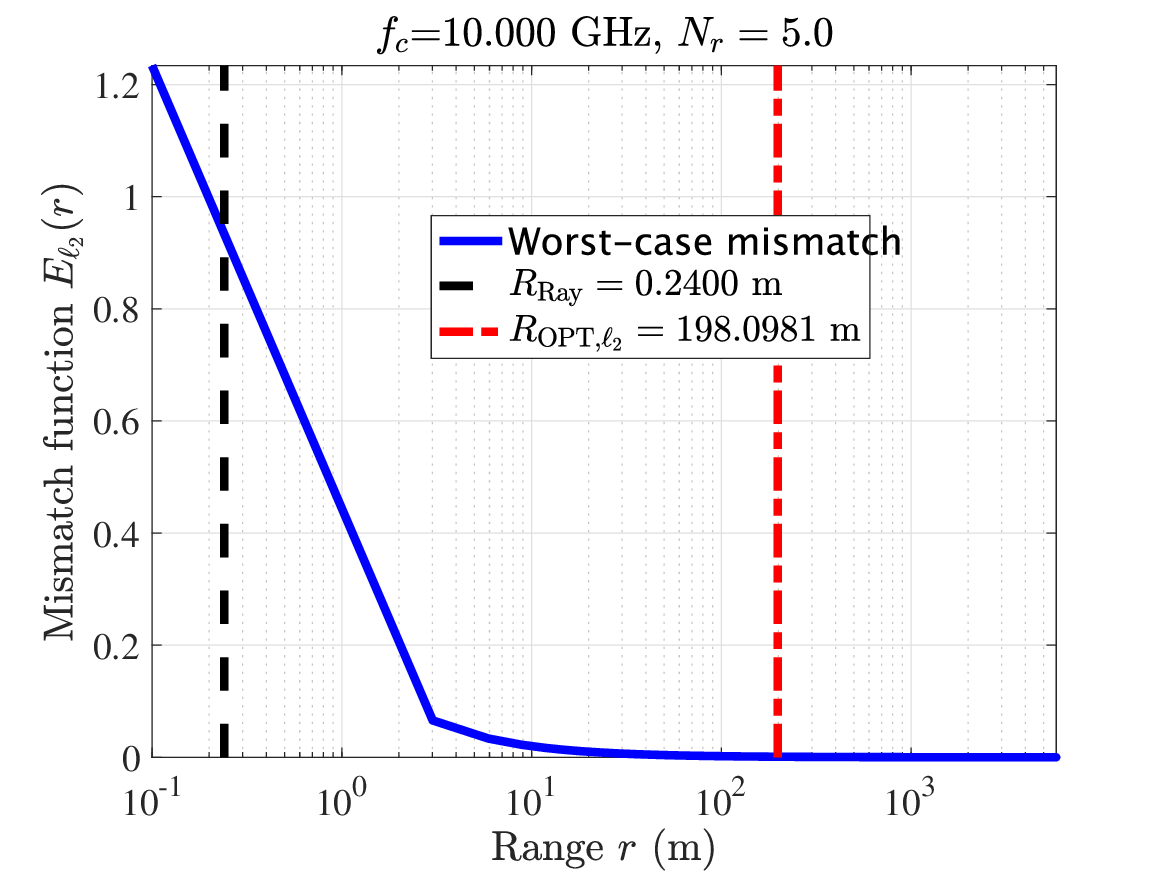}
    \caption{}
    \label{fig:l2_b}
  \end{subfigure}
 \hfill
  \begin{subfigure}[b]{0.32\textwidth}
    \includegraphics[width=\linewidth]{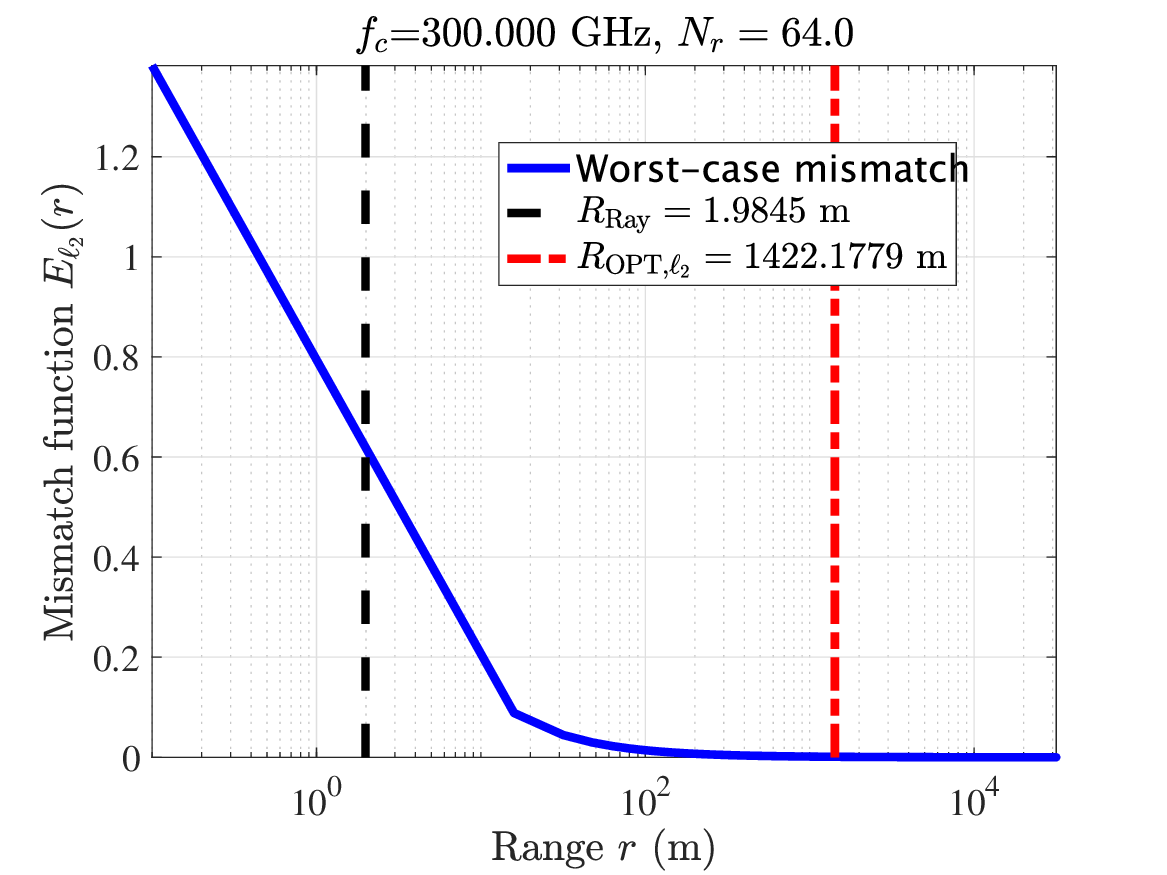}
    \caption{}
    \label{fig:l3_c}
  \end{subfigure}
  \label{fig:linf_l2}
  \caption{(a,b,c) Worst-case element-wise mismatch metric in different $f_c$ and $N_r$. (d,e,f) Worst-case normalized $\ell_2$ mismatch metric.  }
\end{figure*}

\textbf{A. Worst-case element-wise mismatch (\(\ell_\infty\)).}
Figures~\ref{fig:l_infa}-\ref{fig:l_infc} plot
\(E_{\ell_\infty}(r)\) versus range for \(f_c\in\{1, 10, 300\}\) GHz
and \(N_r\in\{2, 5, 64\}\).
The closed-form SPF and SSPF radii closely match the numerically
optimal distance \(R_{\mathrm{OPT},\ell_\infty}\) over the entire parameter sweep,
while the Rayleigh point is up to an order of magnitude smaller.
At 300 GHz and \(N_r=64\) the optimal boundary is \(56.0013\) m,
whereas Rayleigh predicts only \(1.98\) m, illustrating the
dramatically enlarged Fresnel region at sub-THz.

\textbf{B. Normalized \(\ell_2\) mismatch.}
Figures~\ref{fig:l2_a}-\ref{fig:l3_c} examine the
worst-case NMSE metric \(E_{\ell_2}(r)\) for the same carrier set.
With \(\delta_{2}=10^{-3}\) the mismatch remains above tolerance until
ranges far beyond Rayleigh; e.g., at 300 GHz and \(N_r=64\)
the optimal distance is \(1422.18\) m versus the
\(1.9845\) m suggested by Rayleigh. For example, in this specific case, the $\ell_2$ mismatch $E_{\ell_2}(r)$ at the Rayleigh distance is about $0.6$, which leads to the best-case (smallest) NMSE bias floor (see \eqref{eq:NMSE_bias}) of $E_{\ell_2}^2=(0.6)^2=0.36$ (approximately –4.4 dB). This implies that, even with an infinite number of pilot symbols and perfect noise averaging, the optimal estimator cannot reduce the NMSE below –4.4 dB as long as the receiver still models the channel with a planar steering vector at the Rayleigh range. 

\begin{figure*}
    \centering
    
  \begin{subfigure}[b]{0.32\textwidth}
    \includegraphics[width=\linewidth]{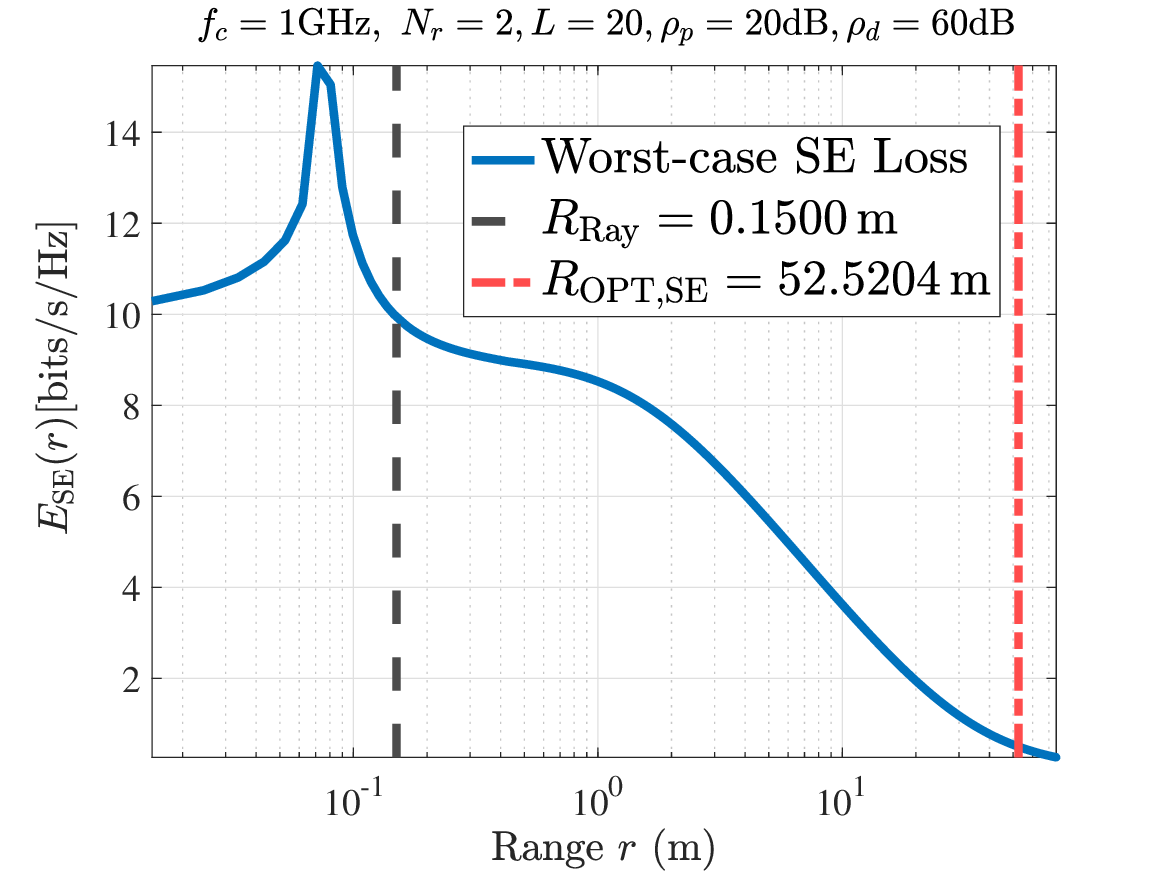}
    \caption{}
    \label{fig:cap_a}
  \end{subfigure}%
  \hfill
  \begin{subfigure}[b]{0.32\textwidth}
    \includegraphics[width=\linewidth]{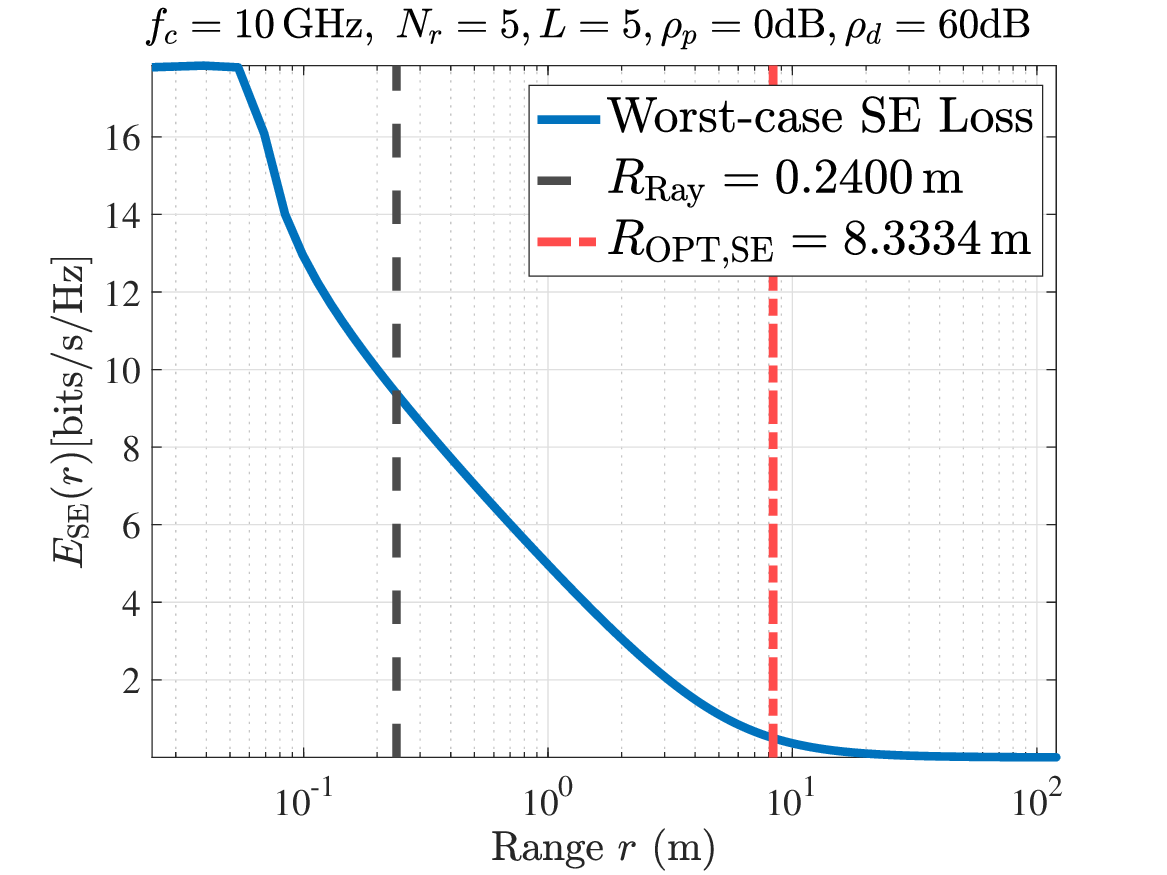}
    \caption{}
    \label{fig:cap_b}
  \end{subfigure}%
  \hfill
  \begin{subfigure}[b]{0.32\textwidth}
    \includegraphics[width=\linewidth]{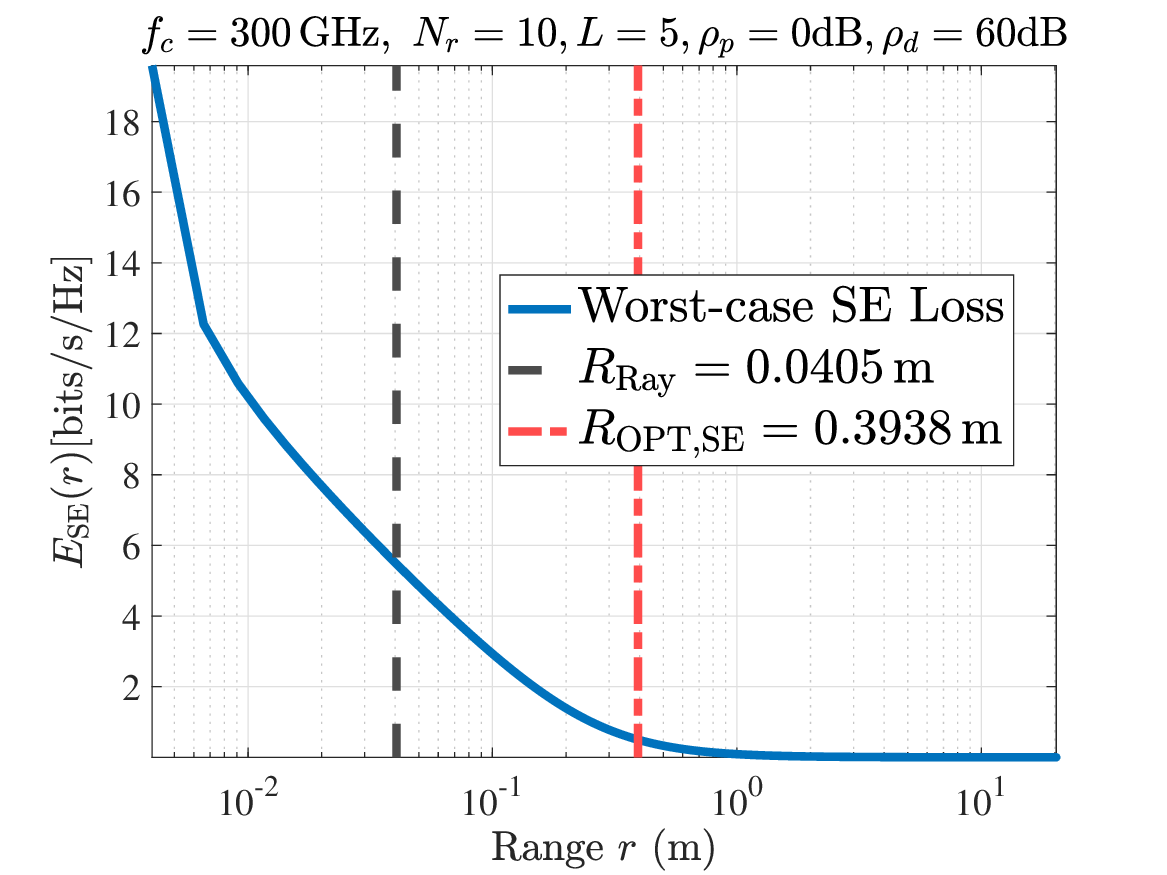}
    \caption{}
    \label{fig:cap_c}
  \end{subfigure}

  \caption{Worst‐case SE loss curves corresponding for varying antenna numbers $N_r$, frequencies and SNR levels. Non-monotonic ripples arise from the oscillatory Fresnel term $\sin(\tfrac{k_{\lambda} D^2}{4r})$ and from angle switching in the worst-case over $\theta$; the SE-safe threshold uses the envelope definition in \eqref{eq:R_opt.se}.}
  \label{fig:capacity}
\end{figure*}

\noindent\textbf{C. Spectral efficiency loss}
Figure~\ref{fig:capacity} depicts the worst-case SE loss
\(\Delta{\rm SE}(r)\) for different carrier frequencies, pilot/Data SNRs and
antenna counts \(N_r=2\), $N_r=5$ and \(10\). The SE loss tolerance is set to $\delta_{\rm SE}=0.5$ bits/s/Hz. In case of $f_c=1 GHz, N_r=2$ shown in Figure \ref{fig:cap_a}, a designer who applies far-field
beamforming right after the Rayleigh range experiences around $10$ bit/s/Hz loss while the maximum possible SE loss is $15$ bit/s/Hz.
At high frequency $f_c=10 GHz$ shown in Figure \ref{fig:cap_b}, the SE loss exactly at the Rayleigh point is around $9$ bits/s/Hz,
confirming that the Fresnel zone must be shifted before planar beams become SE-optimal.
Even in the very high frequency regimes shown in Figure \ref{fig:cap_c}, the SE gap at Rayleigh exceeds \(5\) bits/s/Hz while the optimal threshold is around ten times larger.

\textbf{D. Design take-aways.}
Across the considered metrics and operating points, the proposed thresholds track the optimal distance with high precision: SPF and its strict variant SSPF admit fully closed-form expressions, while EPF--although obtained from a single nonlinear equation--can be evaluated with one simple numeric root-finding approach.
In contrast, the purely geometric Rayleigh rule
systematically underestimates the Fresnel region in high frequencies--often by a full order
of magnitude--leading to premature adoption of planar steering and the
associated losses in array gain, estimation accuracy, and throughput.
These results confirm that \emph{application-driven} boundaries
are indispensable for reliable and resource-efficient 6G deployment.

\section{Conclusions and Future Directions}\label{sec:conclusion}
This paper revisited the fundamental boundary between near- and far-field propagation regimes through the lens of modern 6G systems. Departing from purely geometric criteria, we defined the boundary using three application-driven metrics: worst-case element-wise mismatch ($\ell_{\infty}$), normalized MSE ($\ell_{2}$), and spectral efficiency loss. For each metric, we derived a rigorously optimal transition distance--available either in closed form or via numerical optimization--and validated the results comprehensively across carrier frequencies from 0.1\,GHz to 300\,GHz with arrays of up to 64 elements.

The key insight is that no single universal near-/far-field boundary can simultaneously satisfy all system design layers. Instead, the proposed set $\{R_{\mathrm{OPT},\ell_\infty}, R_{\mathrm{OPT}, \ell_2}, R_{\mathrm{OPT}, {\rm SE}}\}$ provides hardware designers, signal processing engineers, and network planners with clear, targeted benchmarks tailored to specific performance requirements. At millimeter-wave and sub-THz frequencies, these optimal boundaries significantly surpass Rayleigh's classical distance, recovering up to 3\,dB of array gain, eliminating severe NMSE floors, and preventing substantial spectral efficiency losses.

The presented closed-form expressions can be seamlessly integrated into practical engineering workflows, including link-budget planning, calibration procedures, and network optimization frameworks--transforming the near-field region from a potential impairment into a precise and beneficial design parameter. By bridging rigorous Fresnel-region analysis with practical, system-level metrics, this work provides a clear foundation for resource-efficient, reliable near-field communications and sensing, thereby establishing a critical baseline for future 6G deployments. 

The present analysis assumes a static LOS channel and a single-user ULA. In practice, non-LOS components, temporal/channel variations, and multiuser or 2D-array configurations can exacerbate mismatch and shift the effective boundary outward. A pragmatic approach to maintain robustness is to apply calibrated safety margins or probabilistic boundaries based on measured channel statistics. Extending the framework to incorporate such scenarios is left for future work.

\vspace{0cm}
\appendix
\section{Proof of Proposition \ref{prop.optimal_boundary}}\label{proof.prop.optimal_boundary}
In order to make sure that the phase and amplitude mismatch do not exceed a threshold $\delta_{\infty}$, we must have:
\begin{align}\label{eq:rel1}
\max_{\theta\in(0,2\pi)}|\tfrac{1}{R_n} {\rm e}^{-j k_{\lambda} (R_n-r)}-\tfrac{1}{r} {\rm e}^{jk_{\lambda} n d \cos(\theta)}|    \le\delta_{\infty}
\end{align}
By using the relation
\begin{align*}
    &\scalebox{.9}{$|\tfrac{1}{R_n} {\rm e}^{-j k_{\lambda} (R_n-r)}-\tfrac{1}{r} {\rm e}^{jk_{\lambda} n d \cos(\theta)}|\le |\tfrac{1}{R_n}-\tfrac{1}{r}|+\tfrac{1}{r}|1-{\rm e}^{j \Delta \phi(n)}|$}
\end{align*}
a sufficient condition to satisfy \ref{eq:rel1} is the following :

Consider $R_n=\sqrt{r^2+ (nd)^2-2 r n d \cos(\theta)}\triangleq r \sqrt{1+\epsilon_n}$
where we introduced the dimension-free perturbation $\epsilon\triangleq (\tfrac{nd}{r})-2(\tfrac{nd}{r})\cos(\theta)$.
By the Taylor expansion of the function $(1+\epsilon_n)^{-\tfrac{1}{2}}=1-\tfrac{\epsilon_n}{2}+\tfrac{3 \epsilon_n^2}{8}+\mathcal{O}(\epsilon^3)$, we have that
\begin{align}
|\tfrac{1}{R}-\tfrac{1}{r}|=|-\tfrac{\epsilon_n}{2 r} +\tfrac{3\epsilon_n^2}{8 r} +\mathcal{O}(\epsilon_n^3)|    
\end{align}
By replacing $\epsilon_n$ to the above relations, it follows that
\begin{align}\label{eq:amplitude_term}
 |\tfrac{1}{R_n}-\tfrac{1}{r}|\le \tfrac{n d\cos(\theta)}{r^2}-\tfrac{n^2d^2}{2r^3}+\mathcal{O}((\tfrac{nd}{r})^3)   
\end{align}
An angle-free worst-case upper-bound for the above relation is given by $    |\tfrac{1}{R_n}-\tfrac{1}{r}|\le \tfrac{D^2}{2 r^3}
$. For the phase term, we have that
\begin{align}\label{eq:phase_term}
 &\scalebox{.9}{$\tfrac{1}{r}|1-e^{j\Delta\phi(n)}|=\tfrac{2}{r}|\sin( \tfrac{\Delta\phi(n)}{2})|   $}
\end{align}
Thus, if we define the phase mismatch for the \(n\)th element as
\begin{align}\label{eq:phase_err1}
    \Delta\phi(n) \;\triangleq\; k_{\lambda}[(R_n-r) + n\,d\,\cos\theta],
\end{align}
then, the maximum difference becomes in the form of
\[
\Delta\phi_{\max} \;=\; (R_{N_r-1}-r) + D\cos(\theta)\approx \tfrac{D^2}{2r} - \tfrac{D^4}{8r^3} + \tfrac{D^6}{16r^5} + \cdots ,
\]
where we chose the worst-case \(\theta\) such that, \(\sin^2\theta = 1\) in the second term and used Taylor expansion.
For moderately large \(r\) the dominant term is \(\tfrac{D^2}{2r}\).
To enforce the error tolerance $\delta_{\infty}$, we have that $ \tfrac{D^2}{2r^3}+\tfrac{2}{r}|\sin\!(\tfrac{k_{\lambda}\,\Delta\phi_{\max}}{2})| \;\le\; \delta_{\infty},$
which leads to the angle-free strict sufficient condition 
\begin{align}\label{eq:strict_condition}
    \tfrac{D^2}{2r^3}+\tfrac{2}{r}|\sin\!(\tfrac{k_{\lambda}\,D^2}{4r})| \;\le\; \delta_{\infty}
\end{align}
This is a nonlinear condition for the range $r$. We construct the set of ranges $R_{\min}\le r\le R_{\max}$ that satisfies 
$g_{\rm EPF}(r)\triangleq  \tfrac{D^2}{2r^3}+\tfrac{2}{r}|\sin\!(\tfrac{k_{\lambda}\,D^2}{4r})| - \delta_{\infty}  
$
and then we select the first range after the final violation which is $ R_{\rm EPF}\triangleq \sup \{r: g_{\rm EPF}(r)\ge 0\}.$
When the phase term is sufficiently small,
\(\sin x\!\simeq\!x\) with
\(x=\tfrac{k_{\lambda}D^{2}}{4r}\).
Replacing \(\lvert\sin(\cdot)\rvert\) by \(x\) in the exact‐phase
condition~\eqref{eq:strict_condition} gives the \emph{small-angle} inequality $\tfrac{D^{2}}{2r^{3}}
\;+\;
\tfrac{k_{\lambda}D^{2}}{2r^{2}}
\;\le\;\delta_{\infty} $.  Setting the left–hand side equal to the tolerance \(\delta_{\infty}\) gives  
\begin{equation}\label{eq:cubic}
\tfrac{2\delta_{\infty}}{D^{2}}\,r^{3}-k_{\lambda}r-1 = 0 .
\end{equation}
Introduce the depressed–cubic coefficients $p = -\tfrac{k_{\lambda}D^{2}}{2\delta_{\infty}},q = -\tfrac{D^{2}}{2\delta_{\infty}}, 
$. So \eqref{eq:cubic} becomes \(r^{3}+p\,r+q=0\).
For any realistic \(\{D,\lambda,\delta_{\infty}\}\) we have $\Delta = (\tfrac{q}{2})^{2}
        +(\tfrac{p}{3})^{3} < 0$, placing the cubic in the \emph{Casus irreducibilis} regime. By applying the trigonometric solution of Tartaglia/Cardano \cite{cardanoo}, we have: { $R_{\mathrm{SPF}}
=
2\sqrt{-\sfrac{p}{3}}
\cos(
\tfrac{1}{3}\,\arccos(\sfrac{3q}{2p}
\sqrt{-\sfrac{3}{p}})
)$}. The latter equation yields the unique solution
of \eqref{eq:cubic}. Additionally, to find a stricter closed-form for the boundary, we may assume that $r\ge D>1$ and that $ \tfrac{D^2}{2r^3}\le  \tfrac{D}{2r^2}$. This leads to the following stricter condition: $  k_{\lambda}\tfrac{D^2}{2r^2}+\tfrac{D}{2r^2}\le \delta_{\infty}  
$
which leads to the closed-form boundary $R_{\rm SSPF}\triangleq \sqrt{\tfrac{(k_{\lambda}D^2+D)}{2\delta_{\infty}}}$ and the near-to-far condition $r>R_{\rm SSPF}$.

\bibliographystyle{IEEEtran}
\bibliography{refs}

\begin{thebibliography}{10}
\providecommand{\url}[1]{#1}
\csname url@samestyle\endcsname
\providecommand{\newblock}{\relax}
\providecommand{\bibinfo}[2]{#2}
\providecommand{\BIBentrySTDinterwordspacing}{\spaceskip=0pt\relax}
\providecommand{\BIBentryALTinterwordstretchfactor}{4}
\providecommand{\BIBentryALTinterwordspacing}{\spaceskip=\fontdimen2\font plus
\BIBentryALTinterwordstretchfactor\fontdimen3\font minus
  \fontdimen4\font\relax}
\providecommand{\BIBforeignlanguage}[2]{{%
\expandafter\ifx\csname l@#1\endcsname\relax
\typeout{** WARNING: IEEEtran.bst: No hyphenation pattern has been}%
\typeout{** loaded for the language `#1'. Using the pattern for}%
\typeout{** the default language instead.}%
\else
\language=\csname l@#1\endcsname
\fi
#2}}
\providecommand{\BIBdecl}{\relax}
\BIBdecl

\bibitem{rayleigh1896xv}
Rayleigh, ``Xv. on the theory of optical images, with special reference to the
  microscope,'' \emph{The London, Edinburgh, and Dublin Philosophical Magazine
  and Journal of Science}, vol.~42, no. 255, pp. 167--195, 1896.

\bibitem{born2013principles}
M.~Born and E.~Wolf, \emph{Principles of optics: electromagnetic theory of
  propagation, interference and diffraction of light}.\hskip 1em plus 0.5em
  minus 0.4em\relax Elsevier, 2013.

\bibitem{stutzman2012antenna}
W.~L. Stutzman and G.~A. Thiele, \emph{Antenna theory and design}.\hskip 1em
  plus 0.5em minus 0.4em\relax John Wiley \& Sons, 2012.

\bibitem{balanis2016antenna}
C.~A. Balanis, \emph{Antenna theory: analysis and design}.\hskip 1em plus 0.5em
  minus 0.4em\relax John wiley \& sons, 2016.

\bibitem{johnson1984antenna}
R.~C. Johnson and H.~Jasik, ``Antenna engineering handbook,'' \emph{New York},
  1984.

\bibitem{hurd1980ieee}
G.~Hurd, ``Ieee standard test procedures for antennas,'' \emph{Electronics and
  Power}, vol.~26, no.~9, p. 749, 1980.

\bibitem{rappaport2017overview}
T.~S. Rappaport, Y.~Xing, G.~R. MacCartney, A.~F. Molisch, E.~Mellios, and
  J.~Zhang, ``Overview of millimeter wave communications for fifth-generation
  (5g) wireless networks—with a focus on propagation models,'' \emph{IEEE
  Transactions on antennas and propagation}, vol.~65, no.~12, pp. 6213--6230,
  2017.

\bibitem{alkhateeb2014channel}
A.~Alkhateeb, O.~El~Ayach, G.~Leus, and R.~W. Heath, ``Channel estimation and
  hybrid precoding for millimeter wave cellular systems,'' \emph{IEEE journal
  of selected topics in signal processing}, vol.~8, no.~5, pp. 831--846, 2014.

\bibitem{liu2017millimeter}
C.~Liu, M.~Li, S.~V. Hanly, I.~B. Collings, and P.~Whiting, ``Millimeter wave
  beam alignment: Large deviations analysis and design insights,'' \emph{IEEE
  journal on selected areas in communications}, vol.~35, no.~7, pp. 1619--1631,
  2017.

\bibitem{cui2022channel}
M.~Cui and L.~Dai, ``Channel estimation for extremely large-scale mimo:
  Far-field or near-field?'' \emph{IEEE Transactions on Communications},
  vol.~70, no.~4, pp. 2663--2677, 2022.

\bibitem{shahmansoori2017position}
A.~Shahmansoori, G.~E. Garcia, G.~Destino, G.~Seco-Granados, and H.~Wymeersch,
  ``Position and orientation estimation through millimeter-wave {MIMO} in {5G}
  systems,'' \emph{IEEE Transactions on Wireless Communications}, vol.~17,
  no.~3, pp. 1822--1835, 2017.

\bibitem{daei2025near}
S.Daei, A.Zamani, S.Chatterjee, M.Skoglund, and G.Fodor, ``Near-field isac in
  {6G}: Addressing phase nonlinearity via lifted super-resolution,'' in
  \emph{ICASSP 2025-2025 IEEE International Conference on Acoustics, Speech and
  Signal Processing (ICASSP)}.\hskip 1em plus 0.5em minus 0.4em\relax IEEE,
  2025, pp. 1--5.

\bibitem{daei2024timely}
S.~Daei, S.~Razavikia, M.~Skoglund, G.~Fodor, and C.~Fischione, ``Timely and
  painless breakups: Off-the-grid blind message recovery and users’
  demixing,'' \emph{IEEE Transactions on Information Theory}, pp. 1--1, 2025.

\bibitem{razavikia2023off}
S.~Razavikia, S.~Daei, M.~Skoglund, G.~Fodor, and C.~Fischione, ``Off-the-grid
  blind deconvolution and demixing,'' in \emph{GLOBECOM 2023-2023 IEEE Global
  Communications Conference}.\hskip 1em plus 0.5em minus 0.4em\relax IEEE,
  2023, pp. 7604--7610.

\bibitem{daei2025one}
S.~Daei, G.~Fodor, and M.~Skoglund, ``One target, many views: Multi-user fusion
  for collaborative uplink isac,'' \emph{arXiv preprint arXiv:2505.01223},
  2025.

\bibitem{zhang2023near}
X.~Zhang, H.~Zhang, and Y.~C. Eldar, ``Near-field sparse channel representation
  and estimation in {6G} wireless communications,'' \emph{IEEE Transactions on
  Communications}, 2023.

\bibitem{shannon1948mathematical}
C.~E. Shannon, ``A mathematical theory of communication,'' \emph{The Bell
  system technical journal}, vol.~27, no.~3, pp. 379--423, 1948.

\bibitem{molisch2017hybrid}
A.~F. Molisch, V.~V. Ratnam, S.~Han, Z.~Li, S.~L.~H. Nguyen, L.~Li, and
  K.~Haneda, ``Hybrid beamforming for massive {MIMO}: A survey,'' \emph{IEEE
  Communications magazine}, vol.~55, no.~9, pp. 134--141, 2017.

\bibitem{han2015large}
S.~Han, I.~Chih-Lin, Z.~Xu, and C.~Rowell, ``Large-scale antenna systems with
  hybrid analog and digital beamforming for millimeter wave 5g,'' \emph{IEEE
  Communications Magazine}, vol.~53, no.~1, pp. 186--194, 2015.

\bibitem{docomo20165g}
N.~Docomo \emph{et~al.}, ``{5G} channel model for bands up to {100 GHz},''
  Technical report, Tech. Rep., 2016.

\bibitem{lu2014overview}
L.~Lu, G.~Y. Li, A.~L. Swindlehurst, A.~Ashikhmin, and R.~Zhang, ``An overview
  of massive {MIMO}: Benefits and challenges,'' \emph{IEEE journal of selected
  topics in signal processing}, vol.~8, no.~5, pp. 742--758, 2014.

\bibitem{love2005limited}
D.~J. Love and R.~W. Heath, ``Limited feedback unitary precoding for spatial
  multiplexing systems,'' \emph{IEEE Transactions on Information theory},
  vol.~51, no.~8, pp. 2967--2976, 2005.

\bibitem{cardanoo}
R.~W. Nickalls, ``A new approach to solving the cubic: Cardan’s solution
  revealed,'' \emph{The Mathematical Gazette}, vol.~77, no. 480, pp. 354--359,
  1993.

\end{thebibliography}

\end{document}